\documentclass[universe,article,accept,pdftex,moreauthors]{Definitions/mdpi} 
\firstpage{1} 
\pubvolume{1}
\issuenum{1}
\articlenumber{0}
\pubyear{2025}
\copyrightyear{2025}
\externaleditor{Andrea Lapi} 
\datereceived{20 July 2025} 
\daterevised{5 October 2025} 
\dateaccepted{13 October 2025} 
\datepublished{20 October 2025} 
\hreflink{https://doi.org/} 
\usepackage[utf8]{inputenc}
\usepackage[T1]{fontenc}
\usepackage{lmodern}

\Title{Model-Independent Inference of Galaxy Star Formation Histories in the Local Volume}

\TitleCitation{Model-Independent Inference of Galaxy Star Formation Histories in the Local Volume}

\Author{Robin Eappen 
 $^{1,}$*\orcidA{} and Pavel Kroupa $^{1,2,}$*\orcidB{}}

\AuthorNames{Robin Eappen and Pavel Kroupa}

\AuthorCitation{Eappen, R.; Kroupa, P.}

\address{%
$^{1}$ \quad Helmholtz-Institut für Strahlen- und Kernphysik (HISKP), Universität Bonn, Nussallee 14-16, \linebreak 53115 Bonn, Germany\\
$^{2}$ \quad Astronomical Institute, 
 Faculty of Mathematics and Physics, Charles University in Prague, V Holešovickách 2, CZ-180 00 Praha,  
 Czech Republic}

\corres{Correspondence: robin.eappen@gmail.com or robin.eappen@matfyz.cuni.cz 
 (R.E.); \mbox{pkroupa@uni-bonn.de (P.K.)}}

\abstract{Understanding the diversity of star formation histories (SFHs) of galaxies is key to reconstructing their evolutionary paths. Traditional models often assume parametric forms such as delayed-$\tau$ or exponentially declining models, which may not reflect the actual variety of formation processes. We aim to assess what types of SFHs are consistent with the observed present-day star formation rates ($\text{SFR}_0$) and time-averaged star formation rates ($\langle \text{SFR} \rangle$) of galaxies in the Local Volume, without assuming any fixed functional form. We construct a non-parametric framework by generating large ensembles of randomized SFHs for each galaxy in the sample. 
 For each SFH, we compute its predicted stellar mass and present-day SFR and retain only those consistent with the observed values within a 20\% tolerance. We then infer the statistical distribution of power-law slopes $\eta$ (fitted as $\text{SFR}(t) \propto (t - t_{\text{start}})^\eta$) and 50\% stellar mass formation times $t_{50}$. {Across the full sample of 555 galaxies, we find that $\approx$70\% have flat 
 SFHs ($|\eta|\leq0.01$), $\approx$24\% are mildly declining ($\eta<-0.01$), and $\approx$6\% are rising ($\eta>0.01$). In the low-mass bin ($M_\star < 3\times10^{9}\,M_\odot$), rising SFHs slightly increase ($\approx$7\%) but remain a minority as the majority have flat SFHs.} Both $\eta$ and $t_{50}$ correlate strongly with the SFR ratio (Spearman $\rho > 0.75$, $p \ll 10^{-16}$), indicating that the shape and timing of star formation are primarily governed by this ratio. The $t_{50}$ distribution shows sharp spikes near 7.74 and 7.86 Gyr, which we attribute to grid discretization combined with filtering, rather than a physical bimodality. Our results confirm that strongly declining SFH templates are disfavored in the Local Volume: most systems are consistent with flat long-term SFHs, with only mild decline or occasional rising. Importantly, this is demonstrated through a fully model-independent, data-driven approach, with per-galaxy uncertainties quantified using the standard error of $\eta$ and $t_{50}$ from the ensemble of accepted~SFHs.}

\keyword{galaxies; 
 evolution---galaxies; star formation---galaxies; statistics---Local Group methods; statistical methods; numerical} 

\begin{document}
\section{Introduction}
\label{sec:1}
Understanding how galaxies form their stars over cosmic time is a central objective in galaxy evolution studies. A~galaxy’s star formation history (SFH) traces its cumulative growth, encoding the effects of internal processes such as gas accretion, feedback, and~star formation efficiency, as~well as external influences including interactions, environmental quenching, and~cosmological gas supply. Accurately characterizing SFHs is essential for connecting observed properties such as stellar mass, morphology, and~chemical composition to theoretical models of structure~formation.

Traditionally, SFHs are modeled using simple parametric forms such as exponentially declining, delayed-$\tau$, or~lognormal models (e.g., 
 \mbox{\citet{2014ApJS..214...15S, 2013ApJ...762L..15P}}). Among~these, the~delayed-$\tau$ model is commonly adopted in SED-fitting and semi-analytic~\mbox{frameworks}:
\begin{equation}
    \text{SFR}(t) \propto \frac{(t - t_\text{start})}{\tau^2} e^{-(t - t_\text{start})/\tau},
\end{equation}
where $t$ is the time since the begining of the Universe and $t_{start}$ is the time the galaxy began to form its first stars, and $\tau$ is the characteristic time-scale. These models assume that star formation peaks early and declines over time—an assumption often motivated by the observed apparent decrease in the cosmic star formation rate density since $z \approx 2$ (\citet{2014ARA&A..52..415M}). While these forms are mathematically tractable and represent ensemble properties, they may not accurately reflect the full diversity of SFHs, particularly for low-mass, gas-rich, or~isolated systems \mbox{(\citet{Iyer2017}).}

Recent observations of galaxies in the Local Volume (LV)—the nearby $\approx 10$ Mpc region—have provided direct constraints on galaxy-scale star formation. Using tracers such as H$\alpha$ emission and ultraviolet fluxes, current star formation rates ($\text{SFR}_0$) can be measured independently of past histories. In~a comprehensive study, \citet{2020MNRAS.497...37K} compiled a homogeneous catalog of 582 galaxies in the LV, estimating both $\text{SFR}_0$ and time-averaged star formation rates $\langle \text{SFR} \rangle$ over the lifetime of each galaxy. Surprisingly, they found that a large fraction of galaxies exhibit ratios $\text{SFR}_0 / \langle \text{SFR} \rangle \geq 1$, suggesting either constant or even increasing star formation activity until the present epoch. These findings are consistent with earlier studies based on H$\alpha$ imaging and UV SFRs in nearby galaxies (\citet{Karachentsev2013,Hunter2012}).

This result presents a challenge to the delayed-$\tau$ framework, which naturally predicts $\text{SFR}_0 \ll \langle \text{SFR} \rangle$ due to exponential decline. The~findings by \citet{2020MNRAS.497...37K} suggest that galaxies—especially dwarfs and irregulars—may follow flatter or rising SFHs that are inconsistent with common template~assumptions.

To reconcile this, \citet{2023MNRAS.524.3252H} proposed assigning each galaxy an analytically fitted SFH—either delayed-$\tau$ or power-law—such that its observed $\text{SFR}_0$ and inferred stellar mass are reproduced. This approach allowed them to reconstruct a resolved cosmic SFH for the Local Universe and provided important insight into population-level evolution. However, by~design, their method still assigns one specific parametric form per galaxy, limiting the exploration of alternative or degenerate~solutions.

In this work, we develop a complementary, model-independent framework for SFH inference. Instead of assuming a functional form, we generate large ensembles of stochastic SFHs for each galaxy and retain only those histories that simultaneously reproduce the observed $\text{SFR}_0$ and estimated stellar mass $M_\star$ (inferred from $\langle \text{SFR} \rangle$). This filtering process yields a distribution of viable SFHs for each system, reflecting the uncertainty and degeneracy inherent in indirect SFH~reconstruction.

From each accepted SFH, we extract two key~diagnostics:
\begin{itemize}
    \item The SFH slope $\eta$, obtained by fitting a power law of the form $\text{SFR}(t) \propto (t - t_\text{start})^\eta$.
    \item The formation time $t_{50}$, defined as the time when 50\% of the stellar mass has formed.
\end{itemize}

We then analyze how $\eta$ and $t_{50}$ vary as a function of $\text{SFR}_0 / \langle \text{SFR} \rangle$, and~whether galaxies with different star formation activity exhibit systematically different SFH shapes. Our method enables a statistical, non-parametric test of whether rising, flat, or~declining SFHs are permitted by present-day~constraints.

Our results confirm and extend the conclusions of \citet{2020MNRAS.497...37K}: galaxies with high current SFRs relative to their historical average are only compatible with flat or rising SFHs, whereas declining histories are statistically ruled out for most such systems. Importantly, this conclusion emerges not from any assumed SFH form, but~directly from the ensemble of histories allowed by the data. This approach provides a flexible, probabilistic foundation for SFH inference in low-redshift galaxies and sets the stage for future extensions to higher redshift and chemically enriched populations (\citet{Weisz2014,2022ApJ...926..134T}).
\section{Observational~Constraints}
\label{sec:2}
Our analysis is based on the LV galaxy catalogue compiled by \citet{2020MNRAS.497...37K} based on \citet{2004AJ....127.2031K} and \citet{Karachentsev2013}, which includes 582 galaxies within approximately 10 Mpc. This sample was specifically constructed to assess present-day star formation activities in the nearby Universe and provides homogeneously measured estimates of both the current star formation rate ($\text{SFR}_0$) and the time-averaged star formation rate ($\langle \text{SFR} \rangle$) for each~galaxy.

Star formation rates in the LV sample are derived using a combination of far-ultraviolet (FUV) and H$\alpha$, corrected for internal extinction where applicable. This multi-wavelength approach mitigates the systematic biases associated with any individual tracer by capturing both unobscured star formation (traced by FUV and H$\alpha$). The~resulting SFR estimates are thus more robust than those derived from a single indicator, particularly in heterogeneous galaxy populations. The~sample spans a broad range in stellar mass ($10^6$--$10^{10}~M_\odot$), morphological type (including dIrr, and~late-type spirals), and~star formation activity, with~$\text{SFR}_0$ values ranging from $\approx10^{-4}$ to $1~M_\odot~\text{yr}^{-1}$ (\citet{2020MNRAS.497...37K}).

For each galaxy, we compute an estimate of its stellar mass using
\begin{equation}
    M_\star = \frac{\langle \text{SFR} \rangle \cdot t_{\text{sf}}}{\zeta},
\end{equation}
where $t_{\text{sf}} = 12$ Gyr reflects the star-forming lifetime, from~$t_{\text{start}} = 1.8$ Gyr to the present epoch at $t = 13.8$ Gyr, and~$\zeta = 1.3$ is a correction factor accounting for stellar mass loss due to winds and supernovae over~time assuming an invariant canonical IMF (\mbox{\citet{2026enap....2..173K}}, \citet{2025arXiv250906886J}, \citet{2025arXiv250920440G}).

We apply quality cuts by excluding galaxies with unphysical or undefined values—specifically, any with $\text{SFR}_0 \leq 0 \ M_{\odot}/yr$ or $\langle \text{SFR} \rangle \leq 0 \ M_{\odot}/yr$. The~remaining sample constitutes our working dataset, which includes galaxies with active, quiescent, and~intermediate star-forming~states.

To quantify each galaxy’s current activity relative to its historical average, we define the dimensionless SFR ratio as
\begin{equation}
    \mathcal{R} \equiv \frac{\text{SFR}_0}{\langle \text{SFR} \rangle}.
\end{equation}
This parameter serves 
 as the primary control variable in our analysis. The values of $\text{SFR}_0$ and $\langle \text{SFR} \rangle$ adopted here are those compiled by \citet{2020MNRAS.497...37K} for the Local Cosmological Volume sample and are not recalculated in this work. This ensures that the derived ratios \mbox{R = $\text{SFR}_0$}/$\langle \text{SFR} \rangle$ are directly comparable to the observational analysis of \mbox{\citet{2020MNRAS.497...37K}}. A~value of $\mathcal{R} \approx 1$ corresponds to flat SFHs; $\mathcal{R} \gg 1$ implies recent starburst or rising activity, and $\mathcal{R} \ll 1$ indicates a galaxy that formed most of its stellar mass in the past and is now quenched or declining. Across our sample, $\mathcal{R}$ spans more than two orders of magnitude, from~$\approx$0.01 to >10, allowing us to probe a wide spectrum of evolutionary states in a statistically meaningful~way.

{These parameter choices ensure consistency with earlier MOND and IGIMF-based studies: the star-forming timescale $t_{\text{sf}}=12$ Gyr corresponds to the interval from \mbox{$t_{\text{start}}=1.8$~Gyr} (post-reionization onset of sustained star formation) to the present epoch ($t=13.8$~Gyr), while $\zeta=1.3$ accounts for stellar mass loss due to winds and \mbox{supernovae \citep{2020MNRAS.497...37K,2023MNRAS.524.3252H}}. This guarantees direct comparability between our derived $M_\star$, $\langle \mathrm{SFR} \rangle$ and~the values used in these previous works.}
\section{Method: Random SFH Generation and~Filtering}
\label{sec:3}
Our method employs Monte Carlo sampling of stochastic SFHs with rejection based on observational constraints, enabling us to reconstruct the statistically allowed ensemble of SFHs without assuming any functional form. Out of the 582 galaxies in the sample, we retained 555 systems with reliably measured, positive values for $\text{SFR}_0$ and $\langle \mathrm{SFR} \rangle$ ($\text{SFR}_0$ > 0 and $\langle \mathrm{SFR} \rangle$ > 0). The remaining 27 objects, which had non-detections or formally negative SFR estimates, were excluded to avoid undefined or non-physical ($\text{SFR}_0$/$\langle \mathrm{SFR} \rangle$). For~each galaxy, we generate $10^{4}$ randomized SFHs and apply filtering criteria based on its observed star formation rate ($\text{SFR}_0$) and estimated stellar mass ($M_\star$). This process allows us to reconstruct the full range of possible SFH shapes that reproduce a galaxy's current state without relying on any predefined functional~form.

\subsection{Time Grid and SFH~Generation}

We define a uniform time grid between $t_{\text{start}} = 1.8$ Gyr (onset of star formation) and $t_{\text{now}} = 13.8$ Gyr (present cosmic time), discretized into 100 equal steps. {We adopt $t_{\text{start}} = 1.8$~Gyr to represent the onset of sustained star formation after cosmic reionization, consistent with \citet{2020MNRAS.497...37K} and \citet{2023MNRAS.524.3252H}}. At~each step $t_i$, a~random star formation rate is drawn from a truncated Gaussian distribution:
\begin{equation}
    \text{SFR}(t_i) \approx |\mathcal{N}(1, 0.5)|,
\end{equation}
where $\mathcal{N}(\mu,\sigma)$ denotes a Gaussian distribution with mean $\mu=1$ and standard deviation $\sigma=0.5$ (in the dimensionless units of the raw SFH), taking the absolute value then enforces SFR$(t_i)\ge0$ (where the absolute value ensures all star formation rates remain positive). This random sampling permits a wide variety of SFH shapes—from bursty to steady to fluctuating forms—without imposing specific physical models or smoothness~constraints.

Each raw SFH is then normalized so that its time-integrated stellar mass equals unity:
\begin{equation}
    \int_{t_{\text{start}}}^{t_{\text{now}}} \text{SFR}_{\text{norm}}(t) \, dt = 1.
\end{equation}
This normalization ensures that the shape of the SFH can be separated from its amplitude when comparing to galaxy-specific~quantities.

\subsection{Scaling and~Filtering}

To convert the normalized SFH to physical units, we scale it by the estimated stellar mass of the galaxy, computed as described in Section~\ref{sec:2}. The~scaled SFH is given by
\begin{equation}
    \text{SFR}_{\text{scaled}}(t) = \zeta \cdot M_\star \cdot \text{SFR}_{\text{norm}}(t),
\end{equation}
where $\zeta = 1.3$ accounts for stellar mass loss due to winds and supernovae, consistent with previous studies (e.g., \citet{2011ApJ...734...48L, 2014ARA&A..52..415M}).

For each galaxy, we retain only those randomised SFH (out of the $10^{4}$) that simultaneously satisfy the following constraints:
\begin{align}
    |\text{SFR}_0^{\text{model}} - \text{SFR}_0^{\text{obs}}| &< 0.2 \, \text{SFR}_0^{\text{obs}}, \\
    |M_\star^{\text{model}} - M_\star^{\text{obs}}| &< 0.2 \, M_\star^{\text{obs}}.
\end{align}
These acceptance thresholds ensure that the SFHs are tightly anchored to the observed properties while allowing for moderate observational uncertainties and stochastic variability. The~resulting subset of accepted SFHs defines a statistically viable ensemble for each galaxy. (The Catalogue of Neighboring Galaxies \mbox{\citet{2004AJ....127.2031K, Karachentsev2013}} does not list formal measurement uncertainties on either the integrated H\(\alpha\)/FUV SFRs or the stellar masses.  Following Haslbauer~et~al.\ (2024), we therefore adopt a conservative \(\pm20\%\) tolerance on both quantities when filtering our Monte Carlo SFH realizations. {This $\pm20\%$ tolerance corresponds to typical systematic uncertainties in SFR indicators and in stellar mass calibrations, ensuring consistency with the values adopted in \mbox{\citet{2023MNRAS.524.3252H}}}. This choice effectively stands in for the missing catalog errors and encompasses typical calibration and mass‐to‐light–ratio uncertainties at low redshift.)

\subsection{Diagnostic~Quantities}

From each accepted SFH model we derive two quantities that describe its temporal behaviour. 
For a given galaxy, there are $N_{\mathrm{acc}}$ accepted SFHs after the observational filtering step. 
Each SFH is indexed by $i = 1, 2, \dots, N_{\mathrm{acc}}$.

\begin{itemize}

\item \textbf{SFH slope $\boldsymbol{\eta_i}$:} 
For the $i$-th accepted SFH we approximate the star-formation rate as a power law in time, 
\begin{equation}
    \mathrm{SFR}_i(t) \propto (t - t_{\mathrm{start}})^{\eta_i},
\end{equation}
and determine $\eta_i$ via a linear regression in logarithmic space:
\begin{equation}
\log_{10}[\mathrm{SFR}_i(t)] = \eta_i \log_{10}(t - t_{\mathrm{start}}) + \mathrm{const}.
\end{equation}
The slope $\eta_i$ provides a single-parameter measure of the SFH shape: $\eta_i < 0$ corresponds to declining histories, $\eta_i \approx 0$ to approximately constant SFHs, and $\eta_i > 0$ to rising histories.
For each galaxy, we then compute the \emph{median} of the $\eta_i$ distribution,
\begin{equation}
\eta \equiv \mathrm{median}\,(\eta_i),
\end{equation}
which represents the typical SFH slope of that galaxy. 
All figures and correlations in this work use this median value $\eta$ as the galaxy-level diagnostic.

\item \textbf{Formation time $\boldsymbol{t_{50,i}}$:}
For each accepted SFH we define $t_{50,i}$ as the time at which half of the total stellar mass has formed:
\begin{equation}
\int_{t_{\mathrm{start}}}^{t_{50,i}} \mathrm{SFR}_i(t)\,dt 
  = 0.5 \int_{t_{\mathrm{start}}}^{t_{\mathrm{now}}} \mathrm{SFR}_i(t)\,dt.
\end{equation}
This quantity measures how early or late a given SFH assembles its stellar mass. 
Analogously, we take the median over all accepted models for that galaxy,
\begin{equation}
t_{50} \equiv \mathrm{median}\,(t_{50,i}),
\end{equation}
and adopt this $t_{50}$ as the representative half-mass formation time used throughout the~\mbox{analysis}.

\end{itemize}

{It is important to note that $\eta$ is designed to trace the long-term trend of star formation rather than individual burstiness. Short-timescale variability present in high-resolution simulations (e.g. \citet{2016MNRAS.463.3637R}) is averaged out by our fitting procedure, such that $\eta$ represents the overall slope of star formation across cosmic time. For each galaxy, we compute $\eta_i$ and $t_{50,i}$ across all accepted SFHs and adopt the galaxy-level estimates as the medians $\eta \equiv \mathrm{median}(\eta_i)$ and $t_{50} \equiv \mathrm{median}(t_{50,i})$. The associated uncertainties are the \emph{standard errors of the medians}, estimated as $\sigma_{\mathrm{med}} \approx 1.253\,s/\sqrt{N_{\mathrm{acc}}}$ from the same per-galaxy samples (where $s$ is the sample standard deviation and $N_{\mathrm{acc}}$ is the number of accepted SFHs). For binned trends (Section \ref{sec4.3}), means and their standard errors are used instead. Appendix \ref{app:errorbars} figures display error bars as $\pm\sigma_{\mathrm{med}}$.}

The distributions of $\eta$ and $t_{50}$ across the accepted ensemble are then summarized per galaxy, allowing us to analyse population-level correlations and trends with observable quantities such as $\text{SFR}_0 / \langle \text{SFR} \rangle$ and $M_\star$ in the following~sections.

We adopt the median rather than the mean as the representative value for each galaxy because the distributions of $\eta_{i}$ and $t_{50,i}$ obtained from the accepted SFHs are typically non-Gaussian and can exhibit asymmetric tails caused by stochastic filtering. The median provides a robust estimator of the central tendency that is insensitive to outliers or skewed distributions, ensuring that each galaxy’s diagnostic values reflect the statistically most probable rather than the arithmetically averaged SFH behaviour.

\subsection{Correlation~Analysis}
\label{sec:3.4}
To quantify the statistical relationship between the SFR ratio $\mathcal{R} \equiv \text{SFR}_0 / \langle \text{SFR} \rangle$ and the derived SFH properties, we compute both the Spearman rank correlation coefficient and the Pearson linear correlation coefficient. Given two variables $x_i$ and $y_i$ for $i = 1, \dots, N$ galaxies, the~Spearman coefficient $\rho$ is defined as
\begin{equation}
\rho = 1 - \frac{6 \sum_{i=1}^{N} (R_i - S_i)^2}{N(N^2 - 1)},
\end{equation}
where $R_i$ and $S_i$ are the ranks of $x_i$ and $y_i$ in the respective distributions. This non-parametric measure assesses the strength of monotonic association between variables and was originally introduced by \citep{Spearman1904}.

For completeness, we also report the Pearson correlation coefficient $r$, defined as
\begin{equation}
r = \frac{\sum_{i=1}^{N} (x_i - \bar{x})(y_i - \bar{y})}{\sqrt{\sum_{i=1}^{N} (x_i - \bar{x})^2} \sqrt{\sum_{i=1}^{N} (y_i - \bar{y})^2}},
\end{equation}
which captures linear trends between $x$ and $y$ and was developed by \citep{1895RSPS...58..240P}.

In our analysis, we apply these statistics to the pairs $(\mathcal{R}, \eta)$ and $(\mathcal{R}, t_{50})$ across the galaxy sample. The~resulting coefficients and $p$-values are reported in Section~\ref{sec:5}, where we discuss the strength and significance of these trends.
\section{Results}
\label{sec:4}
We now analyze the distribution and correlations of SFH properties derived from the filtered ensemble of accepted histories. For~each galaxy, we compute the median power-law slope $\eta$ and formation time $t_{50}$ across all accepted~SFHs.

\subsection{Distribution of $\eta$ and $t_{50}$}

\begin{figure}[H]
  
    \includegraphics[width=.95\columnwidth]{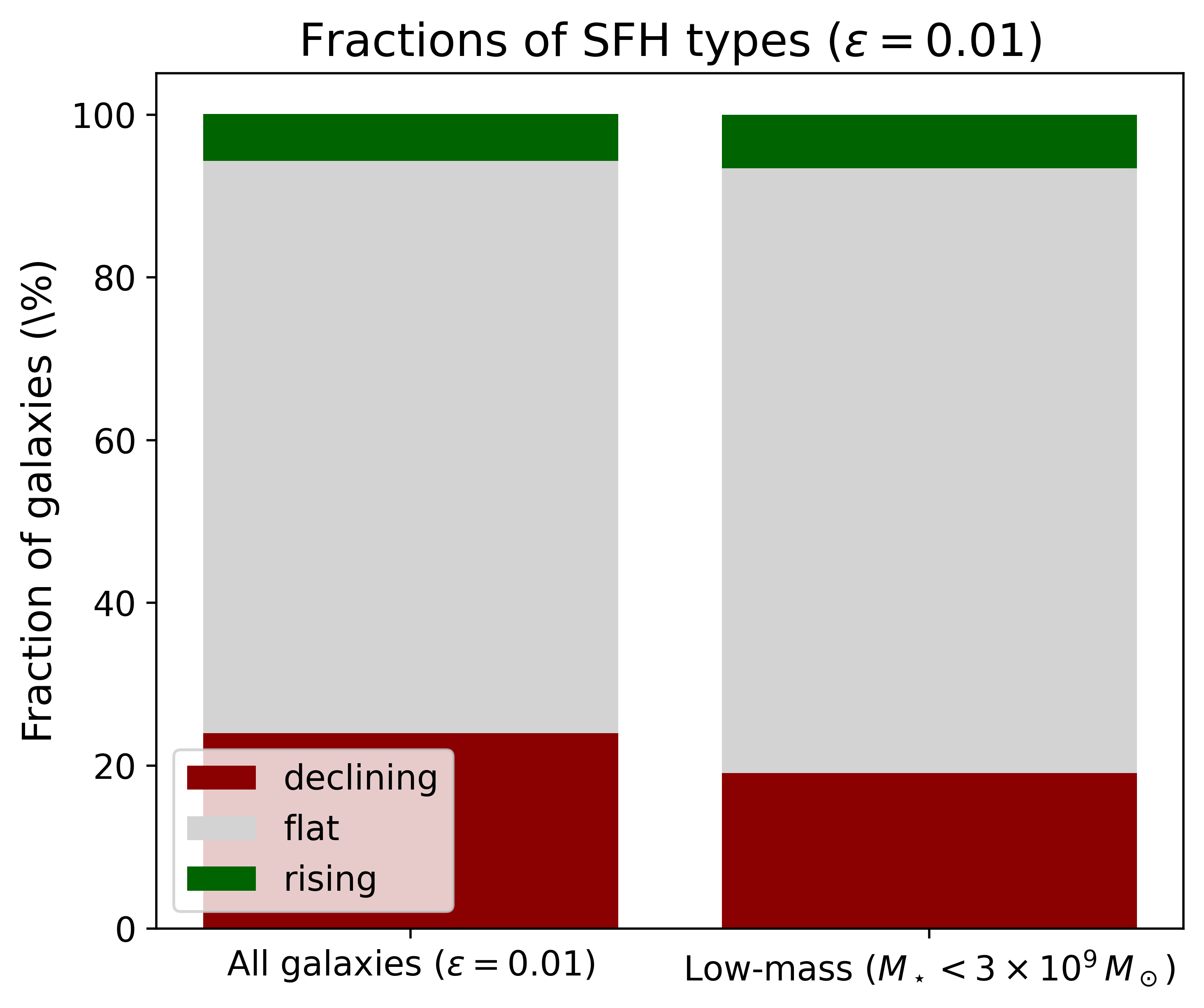}
    \caption{Fraction of galaxies 
 classified as flat, declining, or~rising SFHs based on the median slope $\eta$. Flat is defined as $|\eta|\leq0.01$, declining as  $\eta<-0.01$, and~rising as $\eta>0.01$. These thresholds reflect a tolerance zone that prevents tiny fluctuations from being misclassified as~trends. The left panel is for all galaxies while the right panel is for galaxies with $M_{*} < 3 \times 10^{9} M_{\odot}$}
    \label{fig:eta_barchart}
\end{figure}

We first examine the overall distributions of the two key SFH diagnostic quantities, the power-law slope $\eta$ and the formation time $t_{50}$, computed as the median values across the accepted SFHs for each galaxy. {Quantitatively, across the full sample of galaxies, we find that 
$\approx$70\% are consistent with \textbf{flat} SFHs ($|\eta|\leq0.01$), 
$\approx$24\% are \textbf{declining} ($\eta<-0.01$), and~only 
$\approx$6\% are \textbf{rising} ($\eta>0.01$). 
These fractions confirm that strongly declining histories are rare, and~that extended, nearly flat long-term activity dominates the Local Volume population. 
The threshold of $\pm0.01$ in $\eta$ serves as a tolerance zone: it prevents tiny fluctuations or numerical noise from being misclassified as physical trends. For~the low-mass subsample ($M_\star<3\times10^9\,M_\odot$), the~relative fraction of rising cases is slightly higher ($\approx$7\%), though~the majority ($\approx$74\%) remain flat. 
Figure~\ref{fig:eta_barchart} illustrates the global~\mbox{fractions}.}

\vspace{-6pt}

\begin{figure}[H]
  
    \includegraphics[width=\columnwidth]{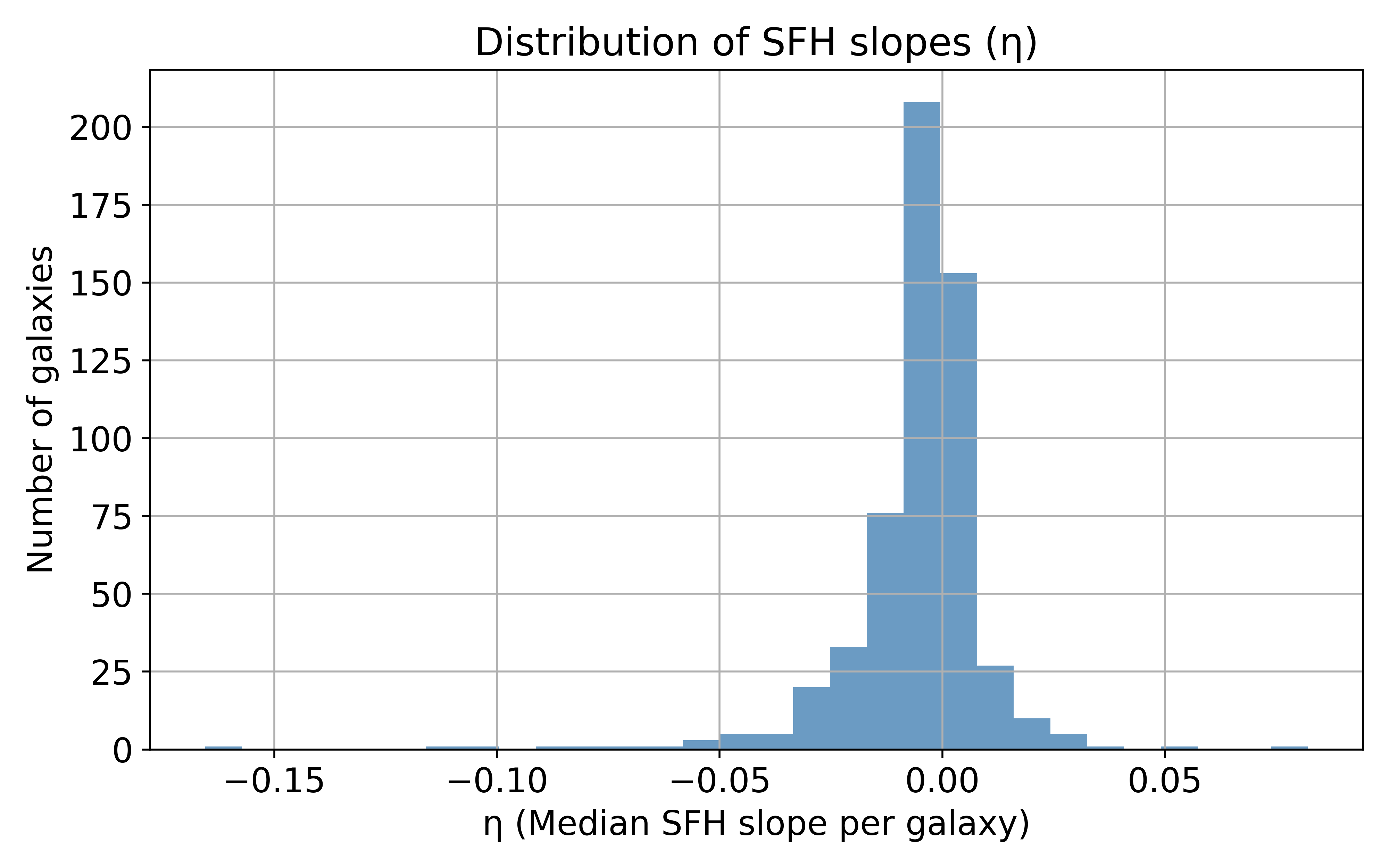}
    \caption{Distribution of median SFH slope $\eta$ across the galaxy sample. The~distribution peaks near $\eta = 0$, indicating that flat SFHs are statistically dominant. A~small tail toward negative $\eta$ suggests that moderately declining SFHs are possible for some galaxies, but~strongly declining or rising histories are~rare. The histogram contains 555 galaxies (see Section \ref{sec:3})}
    \label{fig:hist_eta}
\end{figure}

Figure~\ref{fig:hist_eta} displays the distribution of $\eta$ values. The~majority of galaxies cluster tightly around $\eta \approx 0$, indicating that flat star formation histories are statistically most favored under the observational constraints. The~distribution is sharply peaked and symmetric, with~a full width at half maximum of approximately $\Delta\eta \approx 0.02$ and~a small but significant tail extending to negative values. These tails represent systems with mildly declining SFHs, though~strong declines (\(\eta < -0.05\)) are rare, comprising only a few percent of the sample. Very few galaxies exhibit $\eta > 0.05$, confirming that sharply rising SFHs are uncommon, though~not~excluded.

\vspace{-6pt}

\begin{figure}[H]
    \centering
    \includegraphics[width=\columnwidth]{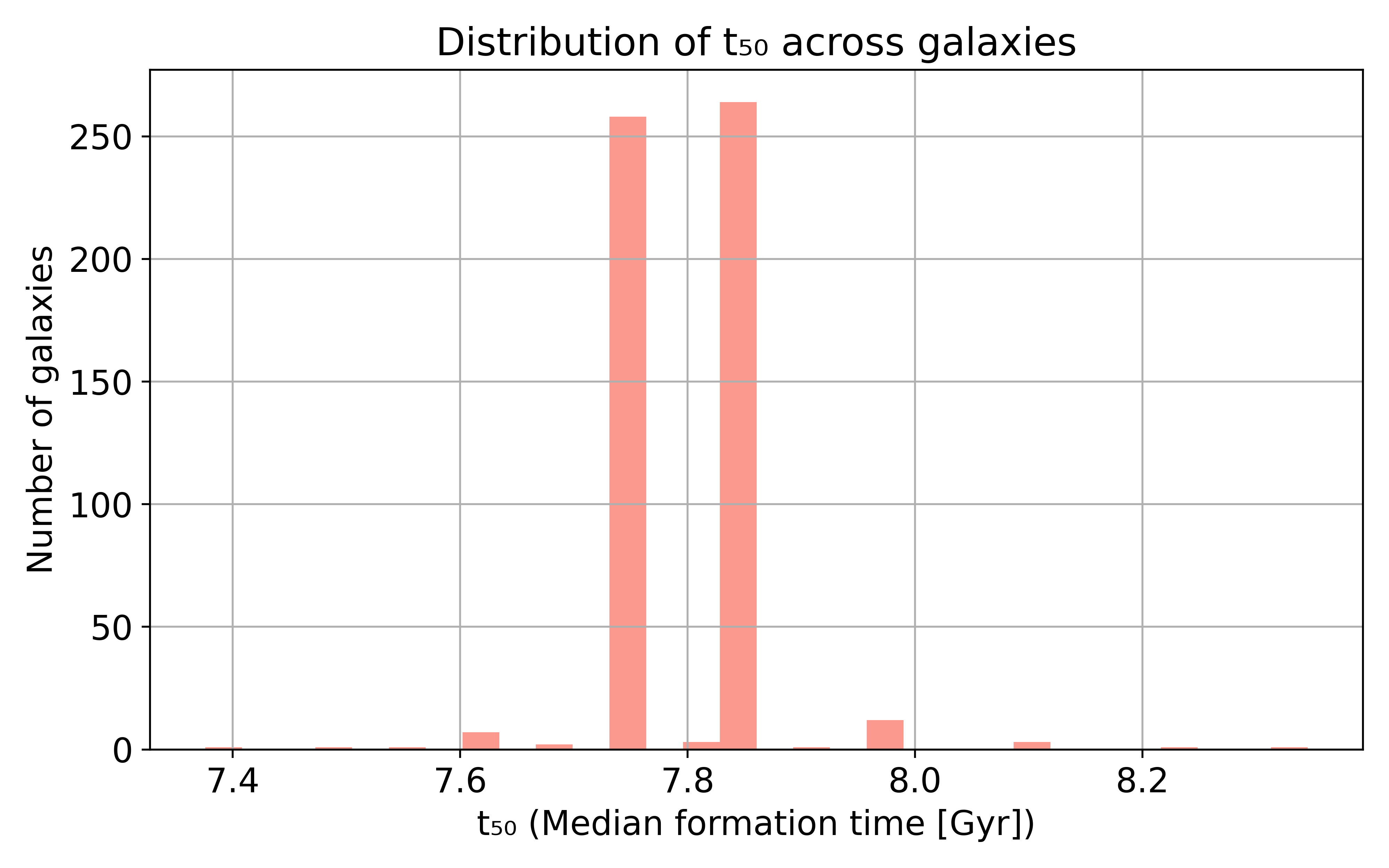}
    \caption{Distribution of $t_{50}$ values across the galaxy sample. The~strong peaks near \mbox{$t_{50}$ $\approx$ 7.72} and 7.86 Gyr indicates that most galaxies formed half of their stellar mass relatively late in cosmic time. The~discrete structure arises from resolution limits in the time grid, but~the concentration toward recent formation is~robust. The histogram contains 555 galaxies (see Section \ref{sec:3}).}
    \label{fig:hist_t50}
\end{figure}

{Figure~\ref{fig:hist_t50} shows the corresponding distribution of $t_{50}$ values—the cosmic time at which 50\% of a galaxy’s stellar mass has formed. Two sharp spikes are visible at 7.72 and 7.86 Gyr. These do not reflect distinct galaxy populations but~instead arise from the finite time grid and the filtering procedure: many accepted SFHs converge on nearly identical half-mass times when constrained by SFR$_0$ and $\langle$SFR$\rangle$. Thus, the~apparent bimodality is a numerical artifact rather than a physical feature. Physically, the~result implies that for the vast majority of galaxies, half of the stellar mass was assembled relatively late—about \mbox{6 Gyr} after the Big~Bang.}

While a few galaxies exhibit $t_{50}$ values extending to 8.3 Gyr, the~absence of earlier\linebreak (e.g., $t_{50} \leq 6$ Gyr) systems suggests that strongly early forming SFHs are statistically disfavored given the observed SFR ratios. This supports the interpretation that most Local Volume galaxies, particularly in the dwarf regime, continue to form stars efficiently at late times and have not experienced early~quenching.

Taken together, these distributions suggest that the filtered SFHs favor extended or slowly evolving star formation, with~little support for rapid early formation or steeply declining activity. This conclusion aligns with the observed SFR ratios presented in \citet{2020MNRAS.497...37K}, and~further confirms that a large fraction of Local Volume galaxies are consistent with prolonged or even sustained stellar mass~growth.

\subsection{Correlations with $\text{SFR}_0 / \langle \text{SFR} \rangle$}

The correlation trends described below are quantified using the Spearman \citep{Spearman1904} and Pearson \citep{1895RSPS...58..240P} coefficients as defined in Section~\ref{sec:3.4}. To~explore how SFH shape and formation timing relate to observable star formation properties, we examine correlations between the diagnostic quantities $\eta$ and $t_{50}$ and the SFR ratio
\begin{equation}
\mathcal{R} \equiv \frac{\text{SFR}_0}{\langle \text{SFR} \rangle},
\end{equation}
\textls[-25]{which serves as a proxy for a galaxy’s current star formation activity relative to its past~average.}

\begin{figure}[H]
    \centering
    \includegraphics[width=\columnwidth]{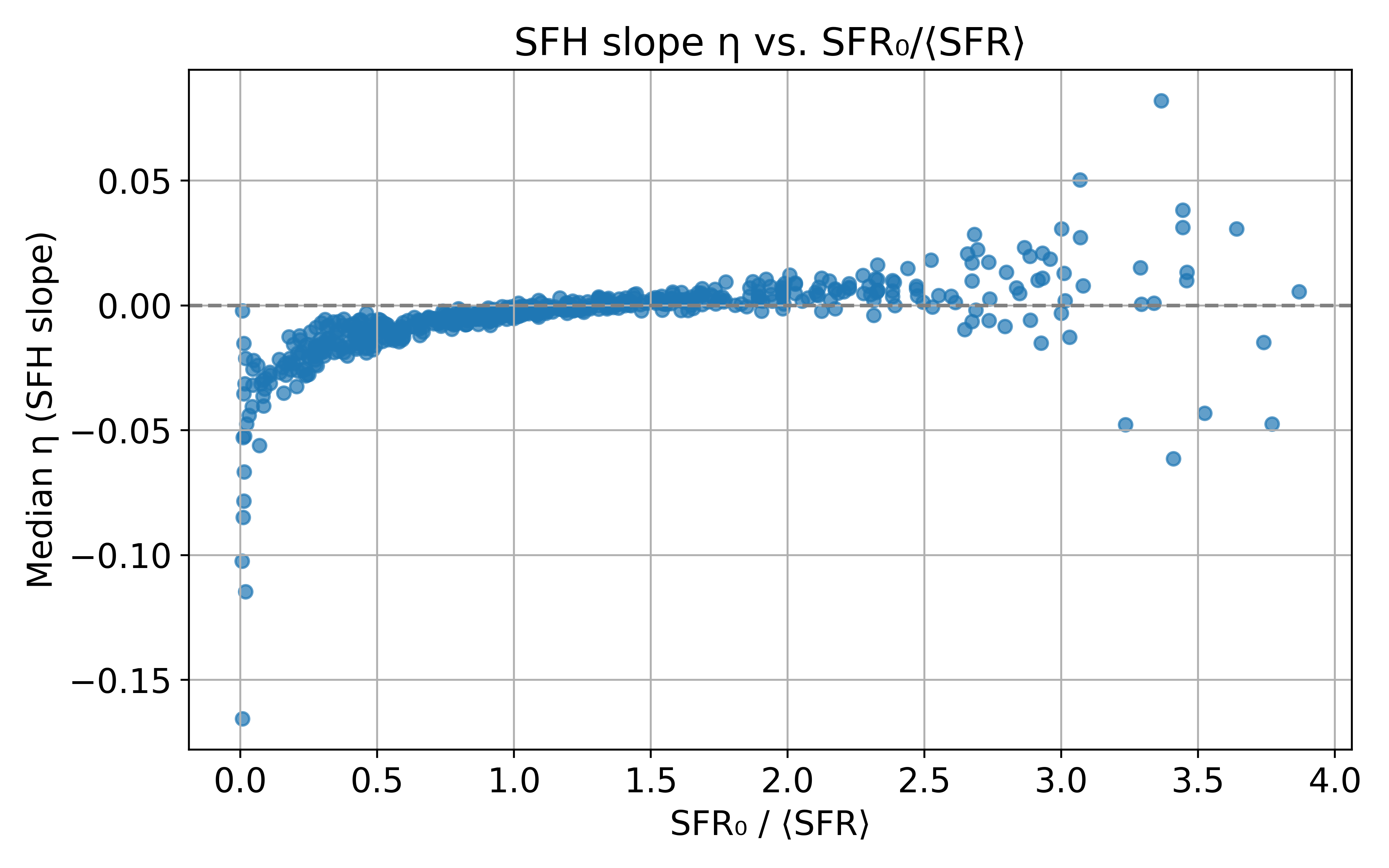}
    \caption{Scatter plot of the 555 median SFH slope $\eta$ for each galaxy versus the observed SFR ratio $R = \mathrm{SFR}_0 / \langle\mathrm{SFR}\rangle$ (from \citet{2020MNRAS.497...37K}). Each point represents the median $\eta$ from the ensemble of accepted SFHs for that galaxy; medians are used because the underlying $\eta$ distributions are often skewed. Galaxies with higher $R$ tend to exhibit rising SFHs.}
    \label{fig:scatter_eta}
\end{figure}

Figure~\ref{fig:scatter_eta} shows a strong and coherent positive correlation between the SFH slope $\eta$ and $\mathcal{R}$. Galaxies with low present-day SFRs compared to their historical average ($\mathcal{R} \leq 0.5$) predominantly exhibit negative values of $\eta$, indicative of declining SFHs. In~contrast, systems with $\mathcal{R} \approx 1$ tend to cluster near $\eta \approx 0$, consistent with flat star formation over cosmic time. Notably, galaxies with $\mathcal{R} > 1$—those forming stars more vigorously now than on average—show progressively more positive $\eta$, with signifying rising SFHs. {A small number of galaxies show negative $\eta$ values even at high $\mathcal{R}$; these are consistent with systems that underwent a recent burst followed by  a rapid decline, or~they may reflect uncertainties in $\mathrm{SFR}_0$ estimates.}

This result directly confirms the expectation that the shape of a galaxy’s SFH must be rising to reconcile high present-day SFRs with a given stellar mass. Since $\eta$ is determined non-parametrically from the filtered ensemble of SFHs, this trend arises statistically rather than by model~assumption.

\vspace{-6pt}

\begin{figure}[H]
    \includegraphics[width=\columnwidth]{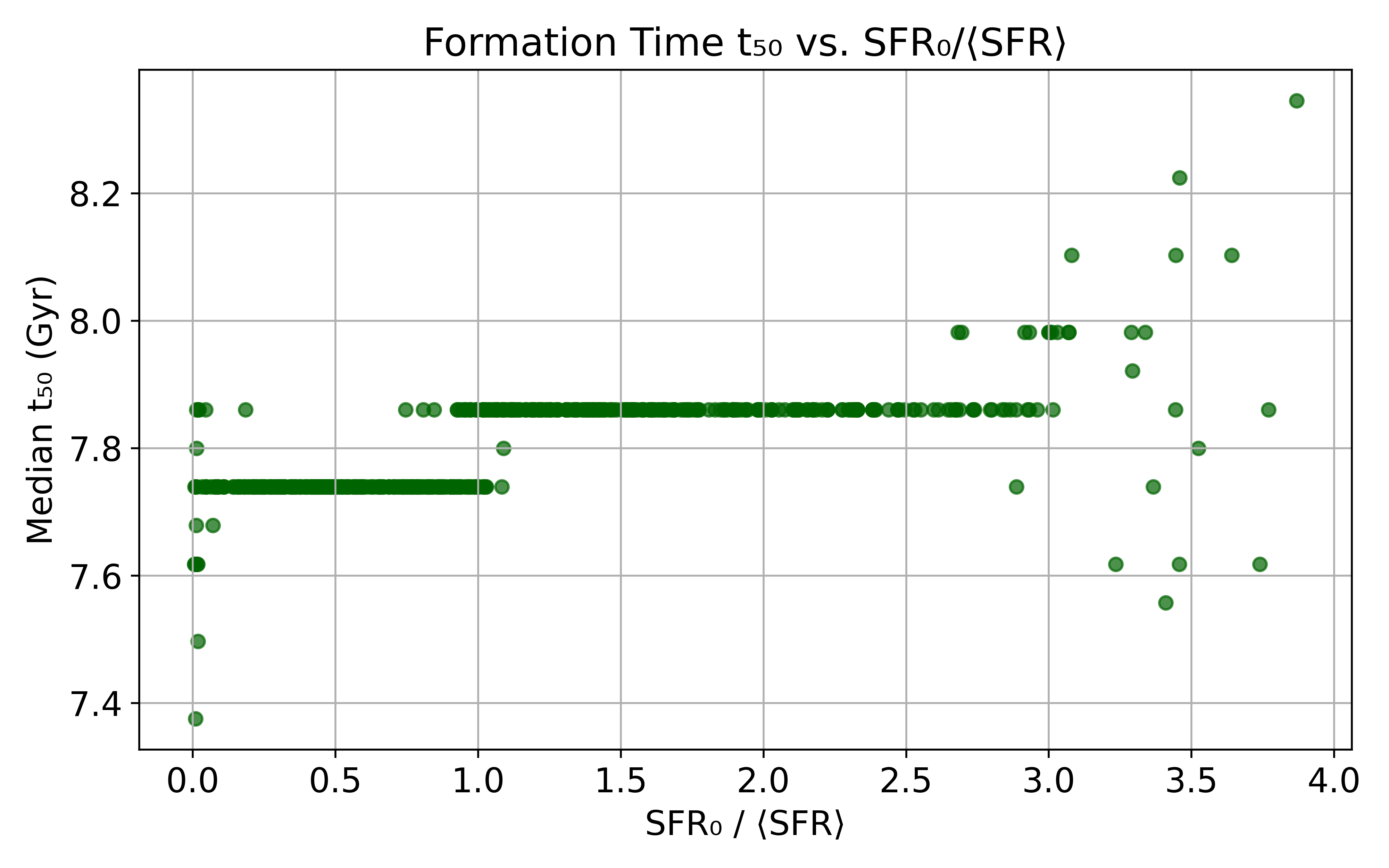}
    \caption{Formation time $t_{50}$ versus $R = \mathrm{SFR}_0 / \langle\mathrm{SFR}\rangle$ (from \citet{2020MNRAS.497...37K}). Points show median $t_{50}$ values per galaxy with $\pm\sigma_{\mathrm{med}}$ (standard error of the median) as vertical error bars; higher-$R$ galaxies form their stellar mass later.}
    \label{fig:scatter_t50}
\end{figure}

Figure~\ref{fig:scatter_t50} displays the corresponding trend between $t_{50}$ and $\mathcal{R}$. A~clear positive correlation is again observed: galaxies with low SFR ratios tend to have earlier formation times ($t_{50}$ $\approx$ 7.6--7.9 Gyr), whereas systems with $\mathcal{R} > 2$ typically show $t_{50} \geq 7.9$ Gyr. In~other words, galaxies that are currently forming stars more actively relative to their past have also formed a larger fraction of their stellar mass more~recently.

Although the $t_{50}$ values are somewhat discretized due to the resolution of the time grid, the~overall trend remains significant and consistent with the $\eta$--$\mathcal{R}$ correlation. Together, these plots demonstrate that both the shape and timing of stellar mass assembly are strongly tied to a galaxy's present-to-past star formation~ratio.

These correlations validate our non‑parametric reconstruction method. Importantly, the~link between $\mathcal{R}$ and both $\eta$ and $t_{50}$ emerges naturally from filtering random SFHs through observational constraints, without~invoking fixed templates or priors. This demonstrates that a galaxy’s current star formation activity carries substantial predictive power regarding its past mass assembly history. The~strength and significance of these correlations are summarized in Table~\ref{tab:correlations}.

\begin{table}[H]
\setlength{\tabcolsep}{15.5mm}
\caption{Correlation coefficients between SFR ratio and SFH~diagnostics.}
\begin{tabular}{lcc}
\toprule
 & \boldmath{$\mathcal{R}$ \textbf{vs.} $\eta$} & \boldmath{$\mathcal{R}$ \textbf{vs.} $t_{50}$} \\
\midrule
Spearman $\rho$ & 0.871 & 0.778 \\
Pearson $r$     & 0.645 & 0.723 \\
$p$-value       & $\ll$$10^{-16}$ & $\ll$$10^{-16}$ \\
\bottomrule
\end{tabular}
\label{tab:correlations}
\end{table}
\unskip

\subsection{Binned~Trends}\label{sec4.3}

While median values are used on a per-galaxy basis to characterize the typical SFH parameters, the binned relations shown below use the mean within each SFR-ratio bin. This choice averages over multiple galaxies per bin to reveal population-level trends, where the distributions are approximately symmetric. Using the mean here smooths statistical noise and allows standard errors of the mean to be displayed as representative uncertainty bars.

To reduce scatter and highlight underlying patterns, we compute binned averages of the SFH slope $\eta$ and the formation time $t_{50}$ as a function of the SFR ratio $\mathcal{R} = \text{SFR}_0 / \langle \text{SFR} \rangle$. Galaxies are grouped into bins of $\mathcal{R}$, and~the mean diagnostic value is computed in each bin, with~error bars representing the standard error of the~mean.

\begin{figure}[H]
    \includegraphics[width=\columnwidth]{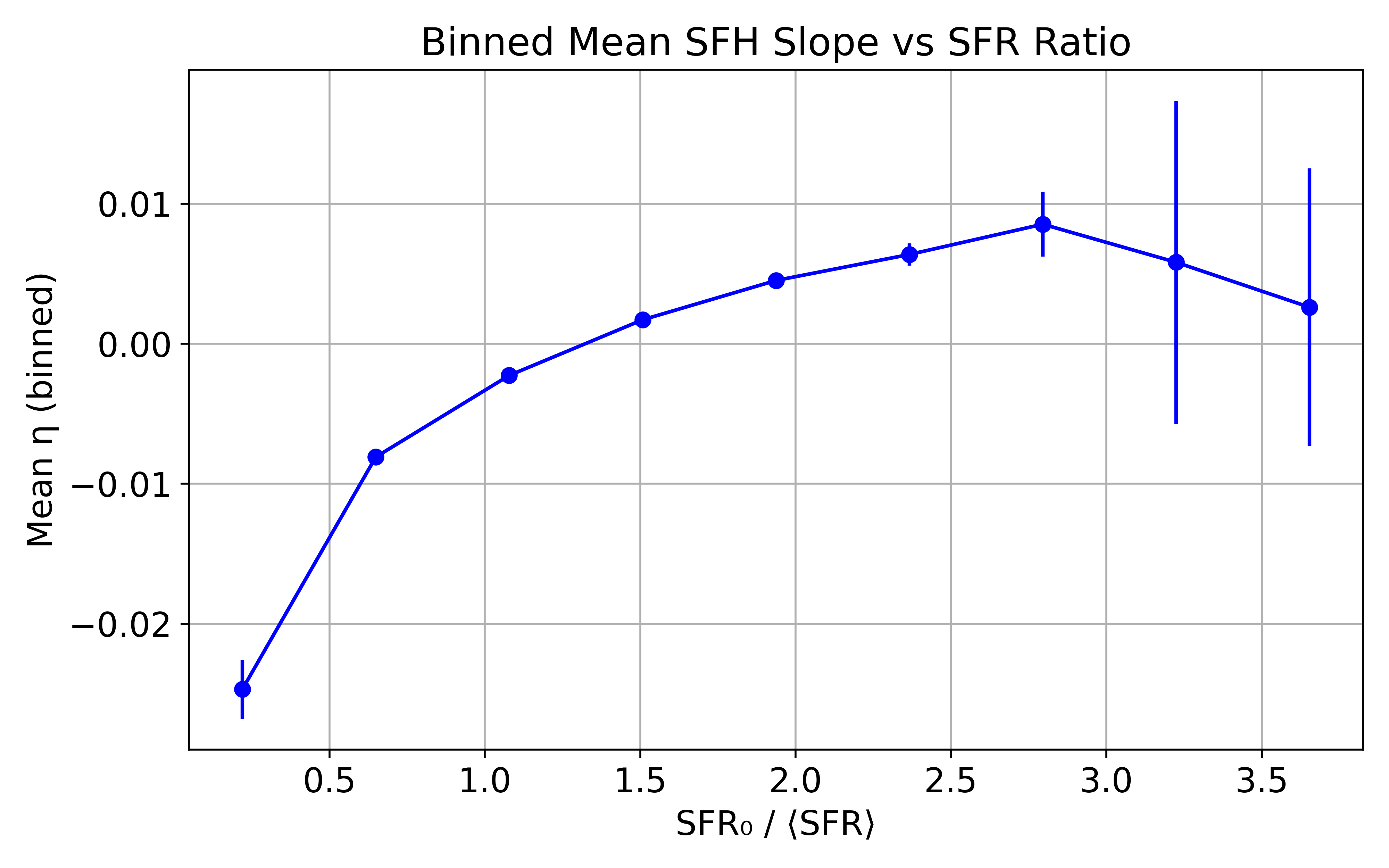}
    \caption{Binned mean SFH slope $\eta$ as a function of the SFR ratio $\mathcal{R} = \mathrm{SFR}_0 / \langle \mathrm{SFR} \rangle$ (values of $\mathrm{SFR}_0$ and $\langle \mathrm{SFR} \rangle$ from \citet{2020MNRAS.497...37K}). Each bin represents the mean $\eta$ of galaxies falling within a given range of $\mathcal{R}$, with vertical error bars showing the standard error of the mean. Galaxies with $\mathcal{R} > 1$ consistently exhibit rising SFHs, 
    while those with $\mathcal{R} < 1$ tend to have flat or declining histories.
}
    \label{fig:binned_eta}
\end{figure}

Figure~\ref{fig:binned_eta} shows the binned trend of $\eta$ as a function of $\mathcal{R}$. The~relation is smooth and monotonic over nearly two orders of magnitude in the SFR ratio. Galaxies with $\mathcal{R} \leq 0.5$ show mildly negative $\eta$ values, indicative of gradually declining SFHs. Systems with $\mathcal{R} \approx 1$ cluster around $\eta \approx 0$, confirming the consistency with flat SFHs, while galaxies with $\mathcal{R} > 1$ exhibit increasingly positive $\eta$, corresponding to statistically rising~SFHs.

This trend confirms that the SFH slope is strongly linked to the relative present-day star formation activity. Importantly, the~absence of significant scatter in this trend suggests that $\eta$ can serve as a robust statistical proxy for $\mathcal{R}$, and~vice~versa.

\begin{figure}[H]
    \includegraphics[width=\columnwidth]{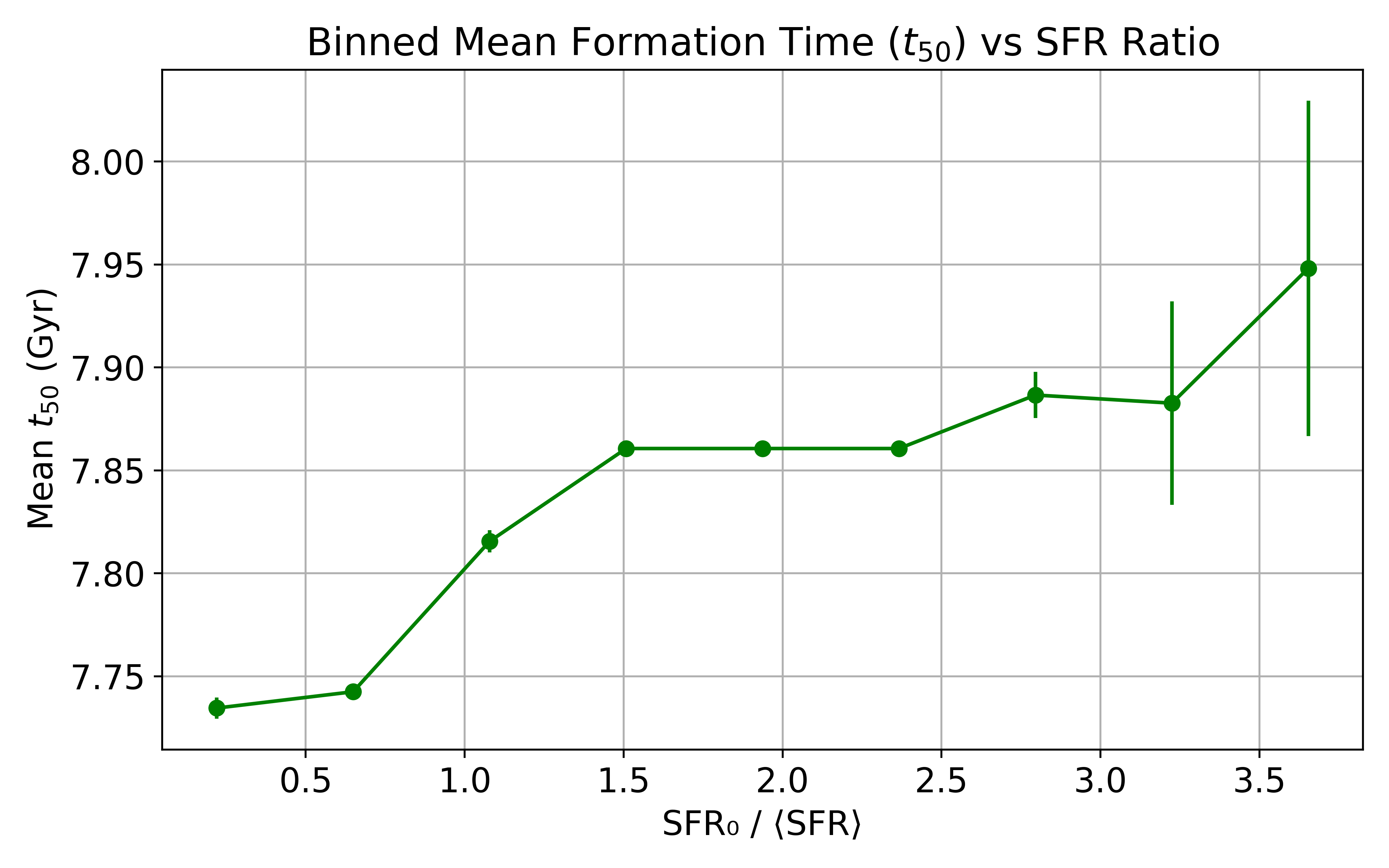}
    \caption{Binned mean formation time $t_{50}$ as a function of the SFR ratio $\mathcal{R} = \mathrm{SFR}_0 / \langle \mathrm{SFR} \rangle$ (from \citet{2020MNRAS.497...37K}). Each bin represents the mean $t_{50}$ of galaxies within a given range of $\mathcal{R}$, with vertical error bars corresponding to the standard error of the mean. Galaxies with higher $\mathcal{R}$ tend to form their stellar mass later in cosmic time, consistent with sustained or rising SFHs.}
    \label{fig:binned_t50}
\end{figure}

Figure~\ref{fig:binned_t50} presents the binned relation between $t_{50}$ and $\mathcal{R}$. A~clear upward trend is observed: galaxies with lower SFR ratios have earlier formation times, while systems with $\mathcal{R} \geq 2$ exhibit $t_{50} \geq 7.86$ Gyr, indicating that a significant fraction of their stellar mass formed at late times. The~final bin ($\mathcal{R} \geq 3.5$) shows the largest mean $t_{50}$, consistent with sustained or even accelerating star formation activity in these~galaxies.

Together with the $\eta$ trend, this result reinforces the interpretation that galaxies with high present-day SFRs relative to their mean are statistically younger in terms of stellar mass assembly. These two diagnostics independently track the shape and timing of star formation, and~both exhibit strong, coherent dependence on $\mathcal{R}$.

The smooth, monotonic nature of both trends across $\mathcal{R}$ suggests that the SFR ratio alone encodes significant information about the recent star formation trajectory of a galaxy. The~consistent scaling of $\eta$ and $t_{50}$ with $\mathcal{R}$ provides a compelling, model-independent confirmation of the physical linkage between a galaxy’s current star formation rate and the statistical shape of its~SFH.

\vspace{-6pt}

\subsection{Dependence on Stellar~Mass}

Finally, we examine whether the inferred SFH slope $\eta$ exhibits any systematic dependence on stellar mass. This is motivated by longstanding observations that more massive galaxies tend to form stars earlier and quench sooner—a trend commonly referred to as “downsizing” (\mbox{\citet{1996AJ....112..839C, 2005ApJ...621..673T, 2019A&A...632A.110Y, 2022MNRAS.516.1081E}}). In~particular, star-forming galaxies have been found to be younger than massive \mbox{ones \citep{2009AIPC.1094..321G}}. If~such effects were present in our sample, we might expect a correlation between $\eta$ \mbox{and the observed $M_\star$.}

\begin{figure}[H]
    \centering
    \includegraphics[width=\columnwidth]{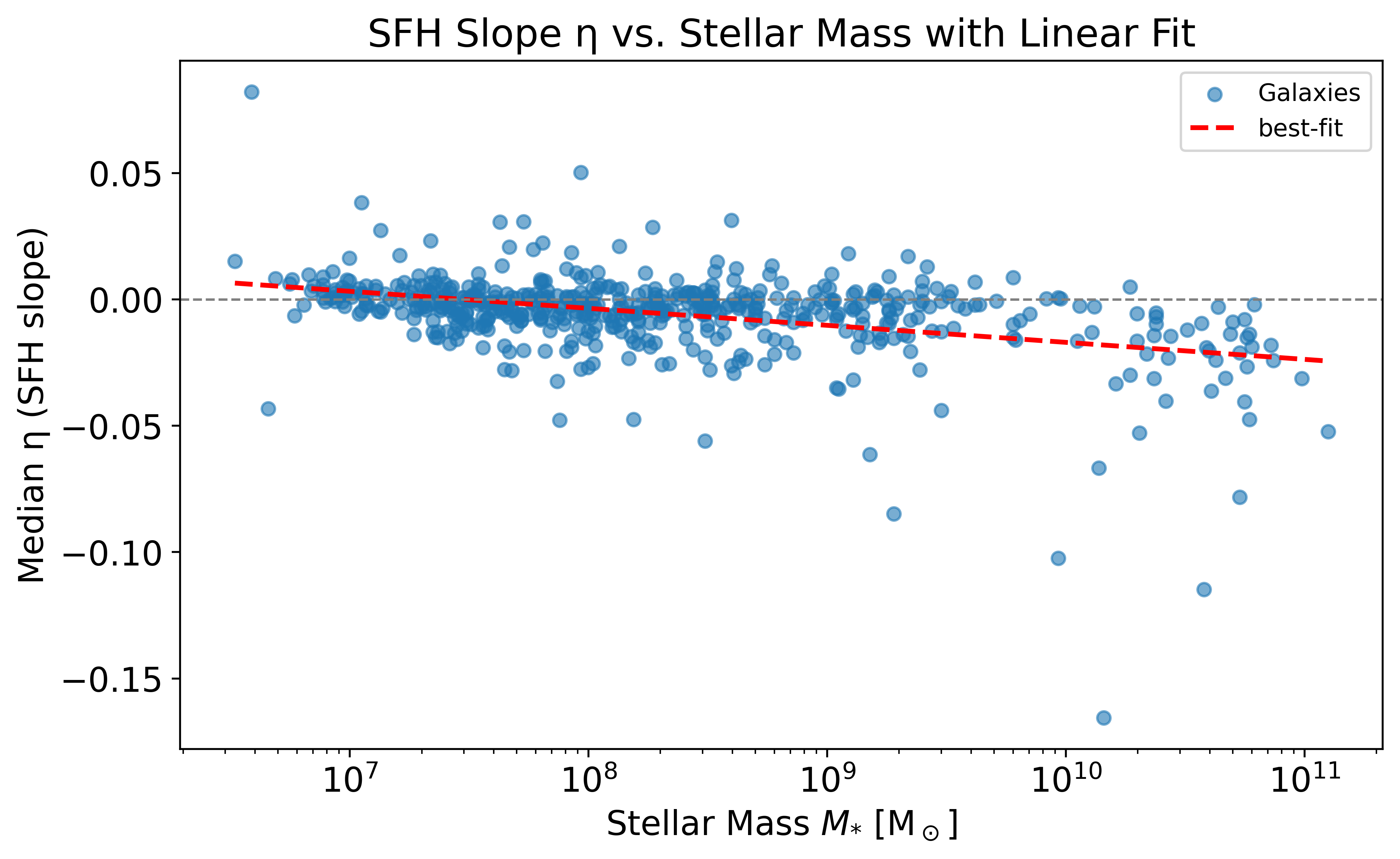}
    \caption{Relation between the median SFH slope $\eta$ and stellar mass $M_\star$. Each point corresponds to one galaxy and shows the median $\eta$ derived from its accepted SFHs. The red dashed line shows the best-fitting linear relation (Equation (\ref{etafit})) in $\log_{10} M_\star$. As in previous figures, medians are used because the $\eta_i$ distributions are often skewed, and $\sigma_{\mathrm{med}}$ provides a robust estimate of the uncertainty on the median.The~slight negative slope indicates a trend whereby more massive galaxies tend to exhibit declining SFHs (results comparable to \citet{2023MNRAS.524.3252H}).}
    \label{fig:eta_mass}
\end{figure}

Figure~\ref{fig:eta_mass} shows the scatter of $\eta$ versus stellar mass for the full galaxy sample. Masses span from $M_\star \approx 10^{7}~M_\odot$ to $10^{11}~M_\odot$ and~are shown on a logarithmic scale. The~blue circles are the galaxies and the red dashed line is the best‐fit relation,
\begin{equation}
\eta_{fit} = (-0.0067 \pm 0.0007)\,\log_{10}(M_\star/M_\odot)\;+\;(0.0503 \pm 0.0055),
\label{etafit}
\end{equation}

This shows a statistically significant decline in the SFH slope with stellar mass. Specifically, the best-fitting relation (Equation (\ref{etafit})) has a slope of $\approx -0.007$, indicating that massive galaxies exhibit declining SFHs ($\eta<0$), while lower-mass systems remain consistent with flat or slowly rising histories ($\eta\approx0$). When divided into mass bins, we find that for $M_\star < 3\times10^{9}\,M_\odot$, approximately 7\% of galaxies are rising, 74\% are flat, and 19\% are declining, as shown in Figure~\ref{fig:eta_barchart}. At higher masses, the distribution skews mildly negative, consistent with more frequent declining SFHs. Although the galaxy-to-galaxy dispersion is appreciable ($\sigma_\eta \approx 0.05$--$0.10$), the correlation remains robust ($r=-0.40$, $p<10^{-20}$), providing quantitative evidence for the downsizing trend in our sample. The increased scatter at the high-mass end likely reflects the greater diversity of SFHs among massive systems. \citet{2023MNRAS.524.3252H} using the resolved-stellar-population catalogue finds a similar pattern: the smallest dwarfs build up their mass late (rising SFR), while the largest galaxies formed early and are now winding down (falling SFR).

This result implies that SFH shape is primarily governed by a galaxy’s relative star formation activity—as captured by the SFR ratio $\mathcal{R} = \text{SFR}_0 / \langle \text{SFR} \rangle$—rather than its absolute scale. The~lack of strong mass dependence suggests that galaxies of similar $\mathcal{R}$ can exhibit comparable SFH slopes regardless of their mass, highlighting the diagnostic utility of $\mathcal{R}$ as a tracer of SFH behavior across diverse~systems. For completeness, we also provide in Appendix~\ref{app:errorbars} the full versions of 
Figures~\ref{fig:scatter_eta}--\ref{fig:eta_mass} with per-galaxy error bars.
\section{Discussion}
\label{sec:5}

Our results provide a statistically grounded, model-independent confirmation of the emerging view that most galaxies in the Local Volume are not well described by classical, exponentially declining star formation histories. Instead, we find that flat and even rising SFHs (for dwarf galaxies) are favored across a large fraction of our sample, particularly for systems exhibiting high $\text{SFR}_0 / \langle \text{SFR} \rangle$ ratios.

This conclusion aligns strongly with the findings of \citet{2020MNRAS.497...37K}, who showed that many nearby galaxies in the Local Cosmological Volume have current SFRs that are comparable to their time-averaged SFRs. Using a sample of 582 galaxies, they argued that declining SFH templates systematically underestimate stellar mass buildup times in such systems and fail to capture ongoing star formation activity. Our results corroborate and extend this claim: by explicitly reconstructing statistically allowed SFHs without assuming any analytic form, we demonstrate that flat SFHs naturally emerge as the only viable solutions. {Independently derived SFHs from color–magnitude diagram (CMD) analyses of resolved nearby dwarf galaxies (\citet{2022MNRAS.513L..40M, 2024ApJ...977...11Y}) similarly show nearly constant SFRs over cosmic time. Although~our method does not use those data directly, the~qualitative agreement between their CMD-based reconstructions and our statistical inference provides an important external consistency check}. {This agreement with CMD-based reconstructions for Local Group dwarfs and the Magellanic Clouds further validates our statistical approach: both methods indicate that flat or gently varying SFHs dominate, with~only a modest minority of rising cases at the lowest masses ($\approx$7\% for $M_\star < 3\times10^9\,M_\odot$).}

\textls[-15]{This stands in contrast to methods that impose specific parametric forms on SFHs—such as exponentially declining, delayed-$\tau$, or~lognormal models—which are commonly used in SED-fitting and semi-analytic models of galaxy evolution (\mbox{e.g., \citet{2014ApJS..214...15S}}}, \citet{2017ApJ...837..170L} but the findings in  \citet{2013ApJ...762L..15P}   imply that the standard approximation of exponentially declining SFHs widely used to interpret observed galaxy spectral energy distributions may not be appropriate to constrain the physical parameters of star-forming galaxies at intermediate redshifts). For~instance, \citet{2023MNRAS.524.3252H} explored the same \mbox{\citet{2020MNRAS.497...37K}} sample but assigned each galaxy a best-fit analytic SFH from a family of power-law and delayed-$\tau$ functions. While that approach recovered similar trends in $\text{SFR}_0 / \langle \text{SFR} \rangle$, it retained the assumption that each galaxy must follow a specific evolutionary trajectory. Our findings show that when no such assumption is made, flat and rising SFHs emerge statistically for a wide range of galaxies—suggesting that these forms are not merely fitting artifacts but inherent to the observational~constraints.

The clearest evidence for this comes from Figures~\ref{fig:scatter_eta} and \ref{fig:scatter_t50} and~their binned counterparts. We find strong, monotonic correlations between the SFR ratio $\mathcal{R}$ and both the SFH slope $\eta$ and the formation time $t_{50}$. These statistical trends, derived using the formal methods presented in Section~\ref{sec:3.4}, are further quantified in Table~\ref{tab:correlations}, which reports both Spearman and Pearson correlation coefficients and confirms the strength and significance of the relationships. Galaxies with $\mathcal{R} \leq 0.5$ predominantly exhibit declining SFHs ($\eta < 0$) and early mass assembly ($t_{50} \leq 7.8$ Gyr), consistent with a quiescent or fading population. In~contrast, galaxies with $\mathcal{R} > 1$ show rising SFHs and late formation times, with \mbox{$t_{50} \geq 7.9$ Gyr} in many cases. This statistically robust correlation supports the interpretation that the current star formation rate relative to the long-term average is a strong predictor of past growth behavior (see also [\citet{2014ARA&A..52..415M}]).

Our findings are also consistent with trends seen in nearby dwarf galaxies, which are thought to often exhibit bursty or stochastic SFHs and do not conform to smooth analytic templates (e.g., \citet{2025OJAp....8E...7T, 2024ApJ...966...25R}). In~such systems, feedback, gas accretion, and~environmental effects probably play important roles in shaping star formation activity, leading to histories that cannot be adequately captured by declining models. The~predominance of flat and rising SFHs in our sample—which is largely composed of low-mass, irregular, and~gas-rich galaxies—supports this physical picture. {High-resolution cosmological simulations (e.g., FIRE, Auriga, NIHAO) predict strongly bursty star formation in dwarfs. Our power-law slope $\eta$  is not intended as a burst diagnostic but instead captures the long-term trend.  In~this sense, our finding that most galaxies are flat or mildly declining is complementary: despite short-term burstiness, their integrated growth remains extended and nearly constant on multi-Gyr timescales.}

Interestingly, we observe no strong dependence of $\eta$ on stellar mass (Figure~\ref{fig:eta_mass}), suggesting that SFH shape is not tightly coupled to a galaxy’s mass. This result contrasts with the commonly observed trend of “downsizing,” in which massive galaxies are thought to form earlier and quench sooner than lower-mass systems (\citet{1996AJ....112..839C, 2005ApJ...621..673T, 2019A&A...632A.110Y, 2022MNRAS.516.1081E}). However, downsizing is primarily evident in quenched, spheroidal populations, whereas our sample is dominated by gas-rich, star-forming systems. In~this context, our finding implies that star formation activity is more closely linked to recent gas supply and regulation processes than to mass alone—consistent with results from \citet{2022ApJ...926..134T}, who show that SFR variations in main-sequence galaxies are better predicted by short-term accretion and feedback behavior than by total stellar~mass.

Methodologically, our approach demonstrates that it is possible to infer meaningful properties of a galaxy’s SFH purely from two observables—$\text{SFR}_0$ and $\langle \text{SFR} \rangle$—without assuming any functional form or relying on stellar population synthesis. This opens the door to new analyses of archival and resolved data using flexible, probabilistic SFH modeling frameworks. For~example, the~same methodology could be extended to higher redshift samples with better SFR estimates (e.g., from~JWST or HST) or combined with resolved CMDs and metallicity gradients to constrain detailed evolutionary~pathways.

The predominance of flat and rising SFHs, combined with late formation times ($t_{50} \approx 7.72$--$7.86$ Gyr) in many low-mass galaxies, presents a potential challenge to predictions from the standard $\Lambda$CDM cosmological framework. In~$\Lambda$CDM-based hydrodynamical simulations, low-mass galaxies are typically expected to form early and rapidly, with~subsequent suppression of star formation due to internal feedback and cosmic \mbox{reionization \citep{2017ARA&A..55..343B}}. These processes produce declining or bursty SFHs, often with $t_{50} \leq 6$ Gyr. However, our model-independent results suggest that many dwarf galaxies in the Local Volume are not ancient relics  but~systems that have continued to build up their stellar mass until recent epochs. The~fact that such trends emerge solely from observational filtering, without~assuming any SFH model, further strengthens the case that early formation and quenching are not ubiquitous among low-mass systems. This discrepancy may point to missing or incorrectly calibrated physics in current $\Lambda$CDM models or~may reflect a fundamental tension with hierarchical formation at small~scales.

In summary, our study confirms that rising and flat SFHs are not outliers, but~the statistically dominant histories permitted by observations of nearby galaxies. These results caution against the default use of declining SFH assumptions in low-redshift studies and underscore the importance of non-parametric modeling in capturing the diversity of galaxy growth. The~strong link between SFR ratio and SFH slope, together with no strong mass dependence, suggests a physically intuitive picture in which galaxies are shaped more by current gas accretion and internal feedback than by total mass—reinforcing the emerging view of galaxy evolution as an extended, rather than impulsive, process.
\section{Conclusions}
\label{sec:6}

In this work, we have developed and applied a non-parametric, statistically grounded framework to reconstruct star formation histories (SFHs) for galaxies in the Local Volume, using only two directly observable quantities per system: the present-day star formation rate ($\text{SFR}_0$) and the time-averaged star formation rate ($\langle \text{SFR} \rangle$). For~each galaxy, we generate $10^4$ randomized SFHs and retain only those that satisfy dual constraints on $\text{SFR}_0$ and stellar mass ($M_\star$), thereby defining an ensemble of viable histories compatible with the~data.

From this filtered ensemble, we extract two key diagnostic quantities: the power-law slope $\eta$, which quantifies the overall SFH shape, and~the formation time $t_{50}$, which indicates when half of the stellar mass was assembled. Our methodology avoids the assumptions of analytic SFH templates, enabling a flexible and physically transparent exploration of what the data actually~permit.

Our main conclusions are as follows:

\begin{itemize}
    \item Galaxies with $\text{SFR}_0 \gtrsim \langle \text{SFR} \rangle$ are statistically consistent only with flat or rising SFHs. Declining SFHs (i.e., $\eta < 0$) are disfavored under the observational constraints. This result confirms and extends previous findings based on parametric models\linebreak (e.g., \citet{2020MNRAS.497...37K}). Low-mass galaxies typically have slightly rising SFRs ($\eta \geq $  0) while massive star-forming galaxies have slightly decreasing SFRs ($\eta \leq$  0), consistent with previous~results.

    \item The SFH slope $\eta$ and the formation time $t_{50}$ both exhibit strong monotonic correlations with the SFR ratio $\mathcal{R} = \text{SFR}_0 / \langle \text{SFR} \rangle$. Galaxies with higher $\mathcal{R}$ tend to have rising SFHs ($\eta > 0$) and later $t_{50}$ values, implying prolonged or accelerating stellar mass assembly. Correlation coefficients exceed $\rho > 0.75$ in both cases (Spearman rank), with~$p \ll 10^{-16}$.

    \item The majority of galaxies have $t_{50} \approx 7.72$–$7.86$ Gyr, suggesting that stellar mass assembly has continued until relatively recent cosmic times. This is especially true for galaxies with high present-day~SFRs.

    \item No strong trend is observed between SFH slope and stellar mass. This implies that SFH shape is not driven by galaxy mass but~is more closely linked to the balance between current and historical star formation activity—potentially reflecting local regulation processes such as gas accretion and~feedback.

\end{itemize}

These findings reinforce the conclusion that standard declining SFH models—such as delayed-$\tau$ or exponential templates—may be inappropriate for describing star-forming galaxies in the nearby universe, particularly in the dwarf and irregular regimes. By~removing assumptions about SFH functional form, our method recovers the full range of evolutionary paths consistent with observational~data.

The framework introduced here can be readily extended. Future studies could incorporate additional constraints such as metallicity evolution, burst indicators, or~resolved stellar population data. It is also well suited for higher-redshift applications, where $\text{SFR}_0$ and $\langle \text{SFR} \rangle$ may be estimated from grism spectroscopy, broadband SED fitting, or~machine learning reconstructions. Overall, our results highlight the power of data-driven SFH inference and provide a new, flexible tool for interpreting galaxy evolution across diverse environments and mass~ranges.

{Statistically, most galaxies in our sample are best described as consistent with flat long-term SFHs: $\approx$70\% fall within $|\eta|\leq0.01$, while $\approx$24\% tilt toward mild decline and only $\approx$6\% toward rising. Thus, although~rising histories are present (especially in the low-mass bin, $\approx$7\% for $M_\star < 3\times10^9\,M_\odot$), the~majority of galaxies cannot be distinguished from flat within the observational uncertainties. This reinforces the conclusion that strongly declining templates are disfavored in the Local Volume, while leaving open the possibility of modestly rising SFHs in some dwarfs—fully consistent with the results obtained by \citet{2020MNRAS.497...37K} and \citet{2023MNRAS.524.3252H}.}
\vspace{6pt} 





\authorcontributions{R.E. performed the calculations, data analysis and first draft of the manuscript. R.E. and P.K. proposed the research idea. R.E. and P.K. developed the manuscript. All authors have read and agreed to the published version of the~manuscript.}

\funding{This work is not part of a funded~project.} 

\dataavailability{No new data were created or analyzed in this study.} 

\acknowledgments{The author R.E. would like to thank KFPP for their support. P.K. would like to thank the DAAD Bonn---Eastern European Exchange program at the University of Bonn and Prague for~support.}

\conflictsofinterest{The authors declare no conflicts of~interest.} 

\appendixtitles{yes} 
\appendixstart
\appendix
\section{Per-Galaxy Error Bars on \boldmath{$\eta$} and \boldmath{$t_{50}$}}
\label{app:errorbars}

\begin{figure}[H]
   
    \includegraphics[width=\columnwidth]{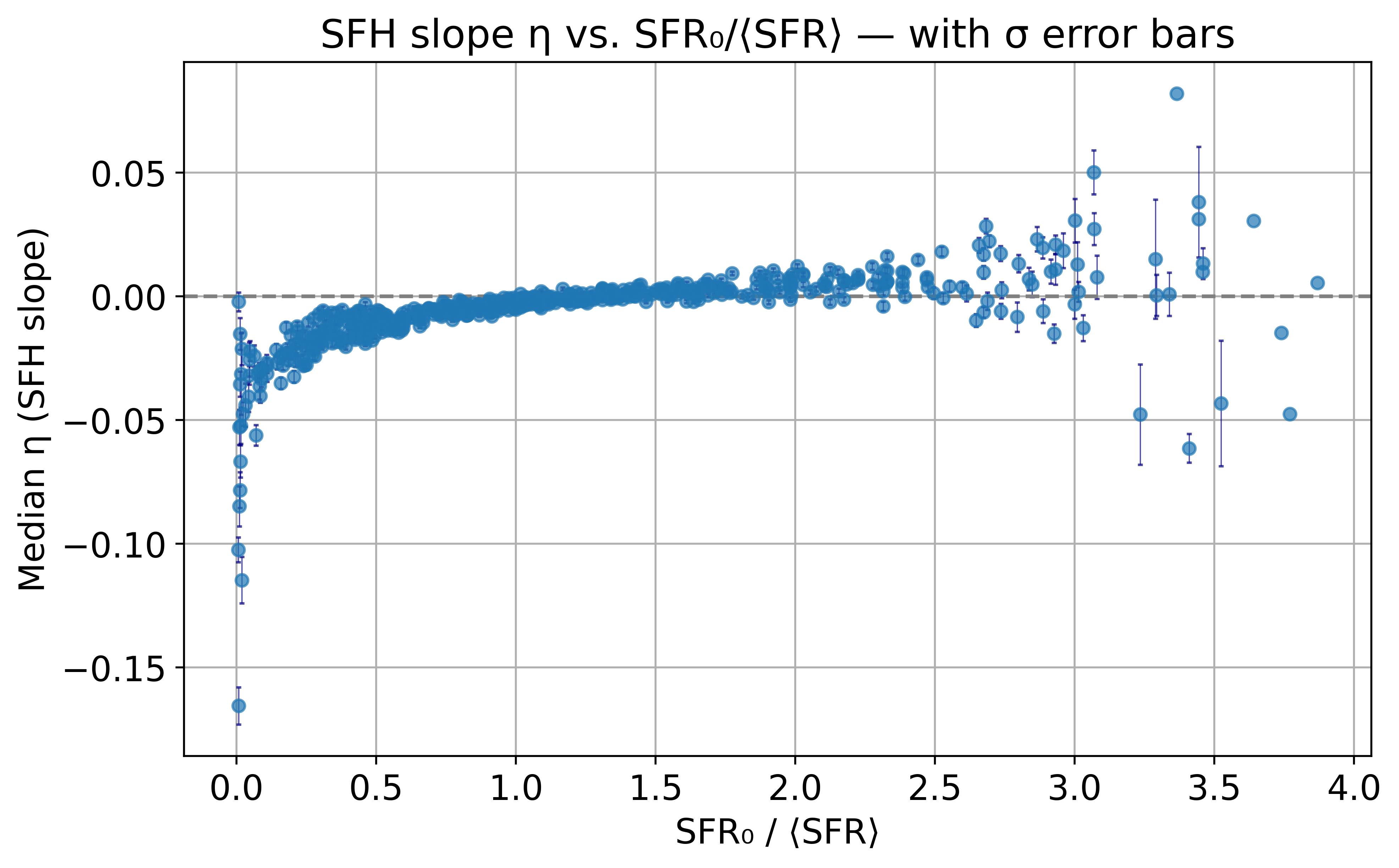}
    \caption{Same as Figure \ref{fig:scatter_eta} with error bars $\pm\sigma_{\mathrm{med}}$ (standard error of the median) computed from the accepted SFHs of each galaxy. The very small error bars in the central region arise because the majority of galaxies exhibit tightly clustered, nearly flat SFHs 
($\eta \approx 0$), yielding a small statistical~\mbox{dispersion}.}
    \label{fig:etavsfrerror}
\end{figure}

\begin{figure}[H]
   
    \includegraphics[width=\columnwidth]{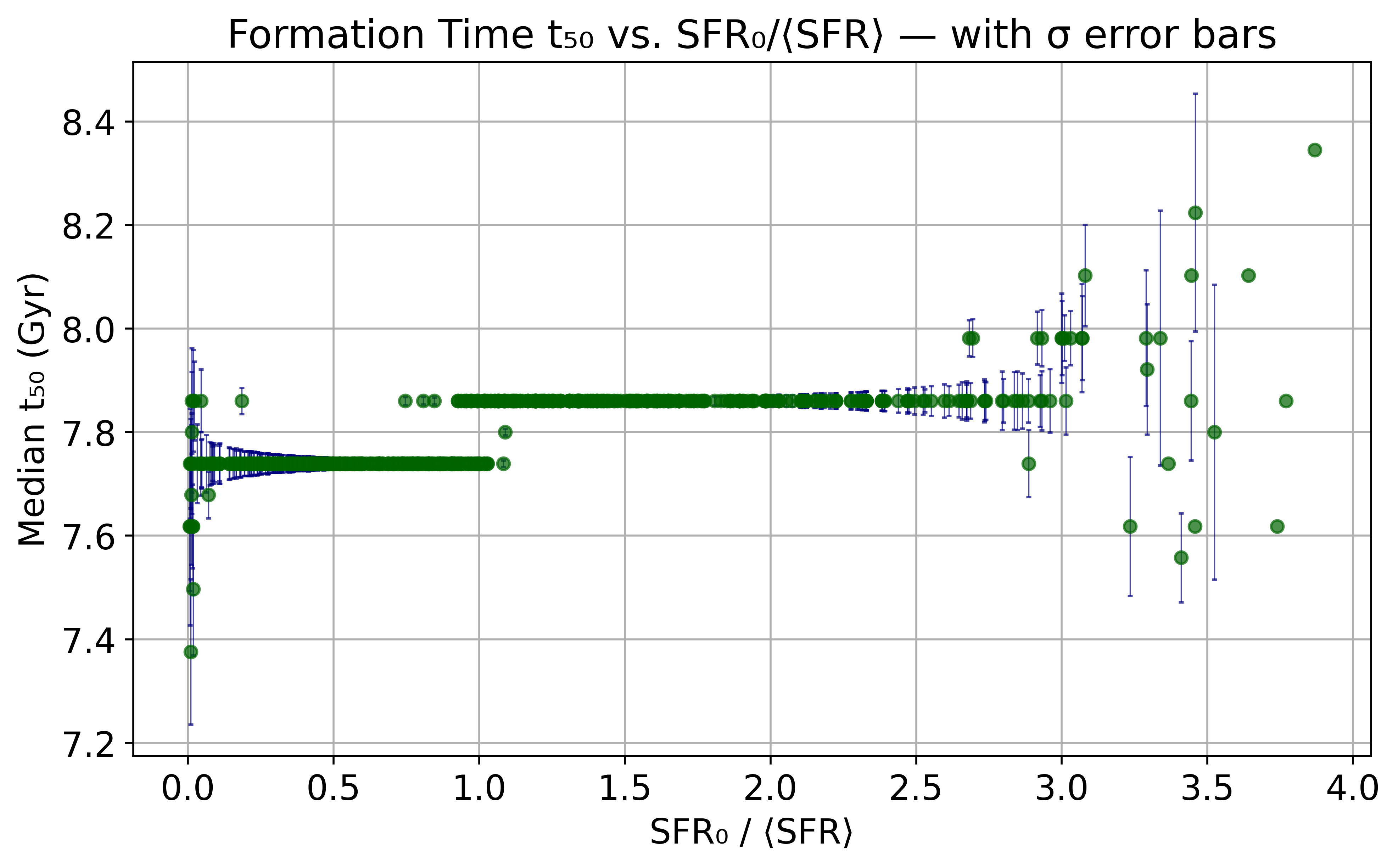}
    \caption{Same as Figure \ref{fig:scatter_t50} with error bars $\pm\sigma_{\mathrm{med}}$ (standard error of the median) computed from the accepted SFHs. The uncertainties are smallest for intermediate $\mathcal{R}$ values where the accepted SFHs converge to similar half-mass formation times ($t_{50} \approx 7.75$–$7.85$ Gyr), indicating robust consistency across the sample.}
    \label{fig:t50sfrrror}
\end{figure}

\begin{figure}[H]
   
    \includegraphics[width=\columnwidth]{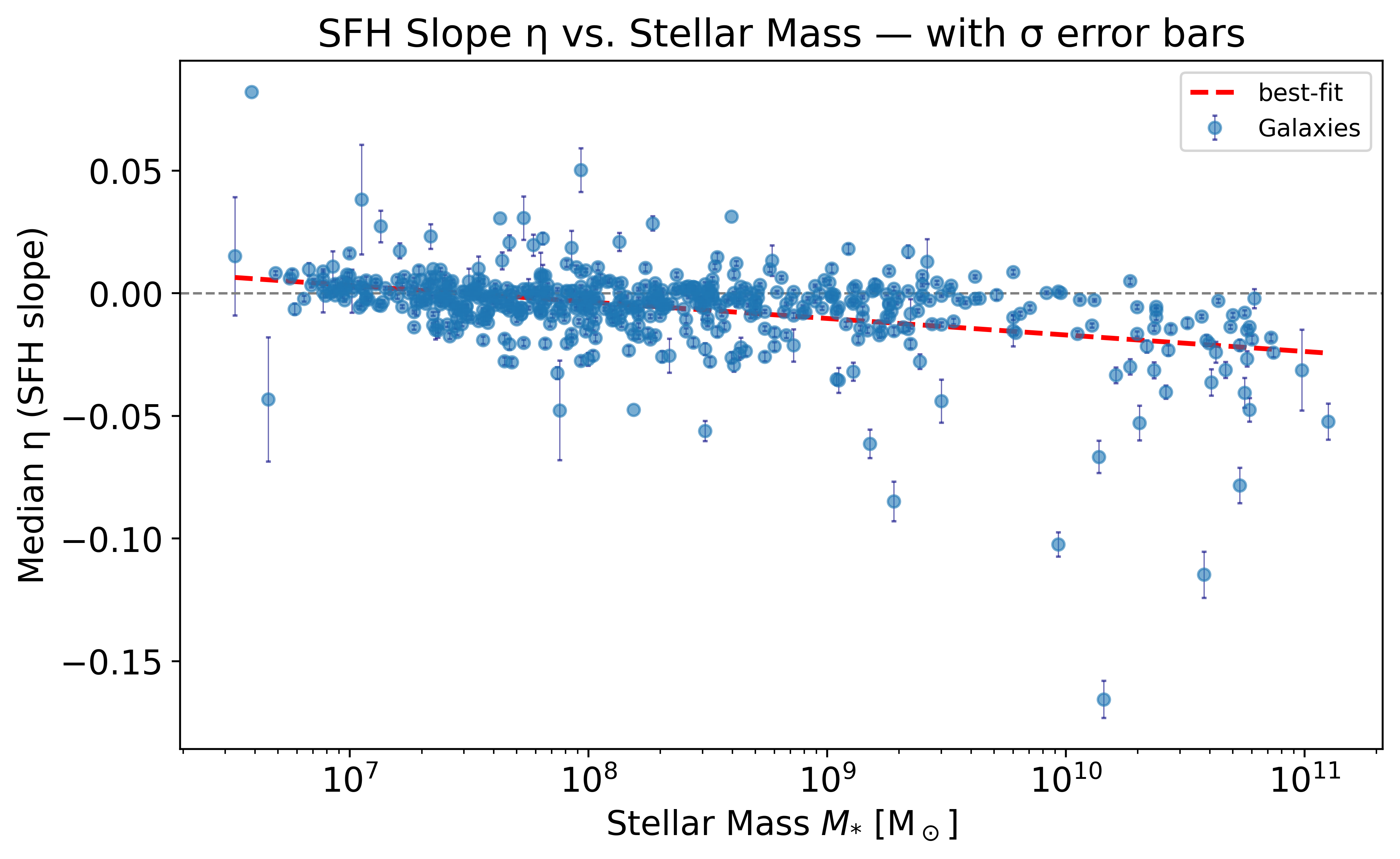}
    \caption{Same as Figure \ref{fig:eta_mass} with error bars $\pm\sigma_{\mathrm{med}}$ (standard error of the median) per galaxy. The small scatter and correspondingly small $\sigma_{\mathrm{med}}$ around intermediate masses reflect that most galaxies exhibit flat SFHs ($\eta \approx 0$), while larger uncertainties at the extremes stem from smaller sample sizes and greater intrinsic diversity.}
    \label{fig:etamasserror}
\end{figure}

Figures~\ref{fig:etavsfrerror}--\ref{fig:etamasserror} present the same relations shown in the main text (Figures~\ref{fig:scatter_eta}--\ref{fig:eta_mass}) but~include per-galaxy uncertainty ranges. For~each galaxy, we compute the distribution of the SFH slope $\eta$ and the half-mass formation time $t_{50}$ across all accepted SFHs from the Monte Carlo ensemble. The~central markers show the median values, and~the vertical bars denote $\pm\sigma_{\mathrm{med}}$ (standard error of the median) computed from the accepted SFHs of each galaxy. This captures the intrinsic stochastic spread arising from the random SFH realizations and from the adopted $\pm20\%$ filtering tolerance on $\text{SFR}_0$ and $M_\star$. The~error bars are omitted from the main figures for visual clarity but are provided here for completeness.

\vspace{-6pt}
\unskip

\begin{adjustwidth}{-\extralength}{0cm}

\reftitle{References}




\PublishersNote{}
\end{adjustwidth}
\end{document}